# Data Mining Application for Cyber Space Users Tendency in Blog Writing: A Case Study

Farhad Soleimanian Gharehchopogh
Computer Engineering Department,
Hacettepe University, Beytepe,
Ankara, Turkey

Seyyed Reza Khaze
Computer Engineering Department, Science and Research Branch, Islamic Azad University,
West Azerbaijan, Iran

## ABSTRACT
Blogs are the recent emerging media which relies on information technology and technological advance. Since the mass media in some less-developed and developing countries are in government service and their policies are developed based on governmental interests, so blogs are provided for ideas and exchanging opinions. In this paper, we highlighted performed simulations from obtained information from 100 users and bloggers in Kohkiloye and Boyer Ahmad Province and using Weka 3.6 tool and c4.5 algorithm by applying decision tree with more than %82 precision for getting future tendency anticipation of users to blogging and using in strategically areas.

## 1. INTRODUCTION
Blog as a recent social media in cyber space is one of the internet and web services [1, 2] which usually provide free software components for users to let them participate as a member of network and virtual community. It also provides unlimited dynamic and interactive relations, and opinions and news about specific issues capable of informing the others opinion about given issues [3].

It causes that this media with its specific capabilities has considerable growth all over the world and especially in Iran [1, 4]. There is a specific proportion between blog features and bloggers tendency with social, political and cultural patterns of different countries and nations [1, 5]. So, recognizing the trend of blogging in the countries such as Iran and different parts of it which has increasing rate, can clarify disadvantages and weaknesses of different countries and the causes related to blogging technological features as well [2, 6, 1 and 5]. This will provide very important strategic programs for different countries and will play an important role in determining strategic policies and planning for social, political, economical and cultural pathologies and providing related strategies.

Recognizing the causes of bloggers tendency and the main parameters of their approach are among major issues which the macro planning of the countries are determined based on these modern technologies and their users and provides vital data for planners and governments. So, it is important to provide proper solutions for determining main factors tends to blogging.

Due to the regional context of the province, population, social and political structure, the users of cyber space in Kohkiloye and Boyer Ahmad in Iran are recognized to each other and their behavior in cyber space will be gathered in a valuable and useful database by cognitive and research methods and through collecting data and analyzing information databases, social networks, blogs, websites and virtual communities which are used by them.

It will help us to survey the causes of the users tends to cyber space, use decision tree method to find out different parameters and get the values by using algorithms and specific patterns [3, 4]. It can be considered to define noted problem in providing a proper solution as well. So far, by using database, decision tree method and analyzing achieved data which are created through statistical methods, we can provide the results that the planners and centers strategically need to determine behavioral policies.

This paper deals with data mining application in cyber space user's tendency in blog wiring. The paper organized as follows. In the next section we review the performed previous works in this domain. Section 3 and sub section of it study the data mining and the process of knowledge finding from data and decision tree algorithm. Section 4 describes the blog wiring ideas and related principles of it in cyber space. In the section 5, we presented a new approach for detecting data mining application in internet based users tendency in blog writing. Finally, the results of this research are considered in section 6.

## 2. PREVIOUS WORKS
In this respect, it can be noted to the articles in the past such as Zafarani et al. [1] are used the Blogizer system and selected proper data and then they began to process it. They try to perceive it by creating lists and extraction of keywords. By measuring the importance and frequency of the gained data, they point out that by automatic programming, social and political issues can perceive.

Nachev and Ganchev [7] are proposed a new approach based on Art2 artificial neural network (ANN) as a kind of data analysis which contributes to pattern survey in data blog and uses it to provide customized content. They filter the data to identify users and meetings and then obtain result vectors and clustering results by using ANN. By analyzing data publish in blogging space, Kwonm et al [8] find out a different theory in contrast with social networks theory which relies on data publish without any relationship. By clustering, they began to find increasing data explosion and correlation between tendency potentials and data explosion clustering [8].

Juffinger and Lex [3] are provided a system for blog analysis in all languages and different subjects by suggesting cross language data survey and imagery tendencies. They believe that the imagery system of tendencies would be based on approaches recognition by providing pre-defined clustering and classifying of blogs. Iraklis Varlamis et al. [9] are considered feature vector after classifying results. By means of analyses techniques, clustering and related vector graphics, integration detection and blog categories distribution along different time intervals for reaching bloggers approach reasons seem possible.

Demartini et al. [5] have used analyses techniques of time intervals in groups and integration method in blogs data to improve ideas on politicians which, at the same time, the estimation of available political trend in blog societies is possible. Wyld [2] has been considered blogging as a social phenomenon while explaining blogging appearance as well and suggested it as unique opportunities to improve interactions and management in digital area.

Then, the author studied partly the statistical survey of bloggers in compared to all internet users and classified in terms of gender, age, race, location and rate of internet access. He also, demonstrated the tendency reasons toward blogging. In following, he noted to the received subjects from bloggers and as a case study, he considered USA blogging.

## 3. DATA MINING

By developing technology and digital media, the data storage technology is provided in data base and resulted in high volumes of them [10, 11]. Due to the high volume, the traditional methods for getting useful methods and suitable patterns from data are difficult and expensive. Sometimes, it was ineffective and didn't recognize potential patterns [11]. Due to the explosive growth of stored data in databases, it was felt the need for new tools to find knowledge and pattern automatically from them.

In the late 1980s, the concept of data mining has been developed which was went further during 1990s [10]. Basically, data mining had been referred to the concept which resulted in useful finding of relations, patterns, tends and potential relationships by using automatic patterns and analyzing the high volume of data and databases which have meaningful and unknown potential patterns [10, 11,12 and13].

Data mining is sophisticated process for recognizing patterns and correct, new and potential models which are placed in high volumes of data in a way. It is understandable for the humans. By the beginning of new millennium, the world scientific media are noted to the data mining as one of the biggest changes of recent decades and the world well-known universities placed it among the 10 superior technologies which resulted in large developments [11].

Data mining has been integrated the relationships among various sciences such as statistics, mathematics, computer, pattern recognition, artificial intelligence and machine learning. It can be achieved to correct anticipation about future by using the noted sciences, processing them and finding pattern to reach self-awareness. These anticipations can be used in many cases such as trading and macro planning of managers to improve and develop the system. By means of data mining, it can be performed data analyses, estimate and anticipate the patterns, classify and cluster the data and perceived association storage [11].

### 3.1 Decision Tree

In artificial intelligence, for better representation and providing clear image of various concepts issues by decision tree, the understanding of audiences will be easier and clearer. In fact, the decision tree is a sample of proper tool and operation for data classification, estimation and providing anticipations due to the features which the data have had [7, 9 and 14]. In fact, the decision tree is more like a real tree which consists of branches, leaves and nodes. The internal nodes are appropriate to specific features and the experiments results of each one would be placed on branches. The leaves nodes are also represent the figure of classes' distributions and in the top stage, the root node is placed [7].

Each node of these leaves revealed a feature and the movement will be started from the root. The determined features are done by this experiment node and based on the results of movement downward. Similarly, this process is repeated by lower grade nodes. The decision tree is an inductive data mining method which continues attributing of data to the first depth or the first width as far as the all data attributed to the specific values of features. In each internal root, the best decision will be made by measuring data irregular values.

There is a set of rules for classifying data to smaller and different groups according to different parameters [8]. The learning and education process of this tree is done by multiple classification classes [14]. The most interested of classification and education algorithms are provided by QUINLAN which includes ID3 and C4.5 algorithms. ID3 algorithm is established by QUINLAN in 1986 which its base was on HUNT algorithm basis and implemented serially [9].

Similar to the other tree educational algorithms, it has also two phases of tree growth and tree pruning. ID3 algorithm uses useful information in classifications and just adopted certain features [8, 14]. C4.5 algorithm is the next generation of ID3 algorithm and uses the type of later pruning law. It is also capable of using discrete features, no-value features and noise data. This algorithm selects the best feature by using irregular criteria and due to applying Gain Ratio factor, it can use many features with high values. If it is no error in educational data, the pruning will be done which cause the tree becomes more general and less- dependent to the educational set.

This algorithm selects a feature by considering the irregularities of each one for the cases which provides to them. After selecting the best feature, no-value features are allocated to the valued features when the data is available and then the algorithm will be continued [7]. Selecting the fact that which features placed in the root will be depending on Information Gain (IG) of each feature. For calculation of IG, it is used formula 1 as follow [5, 16]:

$$IG(S, A) = Entropy(S) - \sum_{v \in Values(A)} \frac{|S_v|}{|S|} Entropy(S_v)$$

**Formula 1-Calculation of the IG**

Which Values (A) are all the features of A, $S_V$ is the subset of S and A has the features of v. Entropy determines the purity rate (irregularity or lack of purity) of the set of examples. If the S set includes the positive and negative examples of a goal concept, S Entropy is defined according to the Formula 2 related to this Boolean category [15]:

$$Entropy(S) = -P_+ \log_2 P_+ - P_- \log_2 P_-$$

**Formula 2- Calculation of the Entropy**

In this formula, $P_+$ and $P_-$ are the proportions of positive and negative examples to total examples, respectively.

## 4. BLOG WRITING

With technological advances and during recent years, web has been considered as a social media and attempted to get it by analyzing to find knowledge, awareness and useful data [1].

Nowadays, due to these advances, blog and blogging societies are constituted important and effective part of world web. So, it can also be attempted to find knowledge discovery process and get information [5].

By using computer networks and widely use of internet, the new terms and expressions are added to the public literature which includes digital writing as well. Digital writing is the process of publishing information as digital data in computer environments which can be facilitated by internet. By publishing these data, the new styles of communication are provided in recent years [1].

Blogs are also a kind of digital writing which the users publish news of newspapers, personal ideas about books and other written media by computers and specially internet rather than personal diary. Due to the wide approaches of humans to internet, it is obvious that their reading habits such as book, newspaper and written media are changed to digital reading. So, blogs become a public media [6]. By using blogs, ones can promote their interactions in cyber space [2].

Due to the wide usage and popularity of internet and human interest to digital reading and writing, blogs are used widely. They also become a tool for groups and different organizations to exploit useful features [4]. As the magazines, books and written media are not bound to a specific one and depend on political and scientific groups and the others in different topics with different incentives, blogs are also followed this by groups and organizations to attract audiences. IT experts, researchers and sociologists are all believe that blogs are social phenomena in cyber space [2]. Due to this unity of thought and as getting public opinions are boring and expensive process through traditional methods, it can be performed easy methods such as polls and surveys to inform social and public ideas [5].

Of course, it would be needed low-cost, easy and concrete methods to discover ideas and prospects through blogs. People use blogs to state their personal and social theories and as they are used by many people, can indicate a good symbol from social space. So, analyzing the blogs are important [3]. As the users' behavior in cyber space is a real appearance of their social behaviors and thoughts, it can perform a true analysis of real space [1]. The studies have shown that personal parameters and features are effective in using blogs and webs and it must be considered unique personal features in analyzing public behavior in webs [17].

As the private satellite networks aren't allowed to work in Iran, major mass media are depend on governmental organizations and institutions and many scientific, social and political changes have been occurred during recent years. So, most people are looking for a place to publish their ideas and opinions and analyze them.

In this respect, they need cheap and simple tool to do them and blogs can support it. Many people use blogs in Iran to publish their thought. Therefore, there are a lot of companies which support free blog facilities. Although these free facilities are provided to attract more users in order to get more income and compensation of free blog offerings. The widespread range of blogs and blogging in Iran, scientific, social and political conditions in blogs data mining and blogger's viewpoint will be resulted in proper and useful analyses of the country.

## 5. CASE STUDY: KOHKILOYE AND BOYER AHMAD PROVINCE IN IRAN

In this paper, we look for to recognize the causes of users tend to cyber space in Kohkiloye and Boyer Ahmad Province in Iran. Collecting information to form database is done by questionnaire. This questionnaire is provided as oral, written and also programming of a website which includes an internet questionnaire and the users can answer the questions as they wish. They entered their used websites, blogs and social networks during the day.

After collecting questionnaires, the wed addresses are gathered to get expected results. And finally, their trustfulness is checked by analyzing their used web pages. As the results were same, for getting better and noiseless response, they will put in database. Better and noiseless responses mean true answers which lead us to better results and promote the precision of decision tree.

We considered the following parameters as questions: age, education, political attitudes, blog topic, and the type of the identity in internet, the influence of managers' inefficiency on tendency, the effect of inefficient media on tendency, the effects of social and political conditions on tendency and finally the effect of poverty in the province on tendency. The noisy or too detailed data in database makes us far from to get proper and suitable answers of algorithms [8]. We pre-processed the data and eliminated some non-relevant data. Finally the followings are considered as the main fields which include: education, political caprice, topics, local media turnover (LMT) and local, political and social space (LPSS). The collected data are shown in Table 1.

In order to get correct answer, we classify bloggers to two groups: professional bloggers and seasonal (temporary) bloggers. Professional bloggers are those who adopt blog as an effective digital media and interested in digital writing in continuous time intervals. Seasonal (temporary) bloggers aren't professional and follow blogging in discrete time periods. In this study, we review the tendency factors considering whether these people are among professional bloggers (Pro Bloggers, PB) and then, consider the other factors according to it.

Due to the performed simulations, we should consider the factors which include the rate and importance of education, the role of political beliefs, interesting topics of users to virtual writing, the effect of state mass media and political and social conditions related to the professional tendency field.

**Table 1- Bloggers database of Kohkiloye and Boyer Ahmad Province in Iran**

| No | Degree | Caprice | Topic | LMT | LPSS | PB |
|---|---|---|---|---|---|---|
| 1 | high | left | Impression | yes | yes | yes |
| 2 | high | left | political | yes | yes | yes |
| 3 | medium | middle | Tourism | yes | yes | yes |
| 4 | high | left | political | yes | yes | yes |
| 5 | medium | middle | News | yes | yes | yes |
| 6 | medium | middle | News | yes | yes | yes |
| 7 | high | left | political | yes | yes | yes |
| 8 | high | right | political | yes | no | yes |
| 9 | high | right | political | yes | no | no |
| 10 | medium | right | Tourism | yes | no | yes |
| 11 | high | right | Tourism | yes | yes | yes |
| 12 | medium | left | News | yes | no | yes |
| 13 | high | left | political | yes | yes | no |
| 14 | low | right | news | no | yes | no |
| 15 | high | left | political | yes | yes | yes |
| 16 | medium | left | impression | yes | yes | yes |
| 17 | medium | left | political | yes | yes | yes |
| 18 | high | right | political | yes | yes | yes |
| 19 | medium | left | impression | yes | yes | yes |
| 20 | high | right | tourism | yes | yes | no |
| 21 | high | left | political | yes | yes | yes |
| 22 | medium | left | news | yes | yes | yes |
| 23 | high | right | political | no | yes | no |
| 24 | low | left | tourism | yes | no | no |
| 25 | high | left | news | yes | yes | yes |
| 26 | high | left | political | yes | yes | yes |
| 27 | low | right | impression | no | no | yes |
| 28 | high | right | political | yes | yes | yes |
| 29 | high | left | impression | no | no | yes |
| 30 | medium | left | scientific | yes | yes | no |
| 31 | high | right | political | yes | yes | yes |
| 32 | low | left | scientific | yes | yes | no |
| 33 | medium | right | tourism | yes | yes | no |
| 34 | Low | right | political | yes | yes | yes |
| 35 | High | left | impression | yes | no | yes |
| 36 | medium | left | tourism | yes | no | yes |
| 37 | medium | middle | scientific | yes | no | yes |
| 38 | medium | middle | impression | no | yes | no |
| 39 | medium | right | scientific | yes | yes | no |
| 40 | medium | left | impression | no | no | yes |
| 41 | High | left | political | yes | yes | no |
| 42 | medium | left | news | no | yes | yes |
| 43 | High | left | political | yes | yes | yes |
| 44 | medium | right | news | yes | yes | no |
| 45 | medium | left | tourism | yes | no | yes |
| 46 | medium | middle | news | yes | yes | yes |
| 47 | Low | middle | impression | yes | no | no |
| 48 | Low | right | impression | yes | no | no |
| 49 | medium | right | news | yes | yes | no |
| 50 | medium | left | impression | yes | yes | yes |
| 51 | High | left | political | yes | yes | yes |
| 52 | High | left | political | yes | yes | yes |
| 53 | medium | middle | tourism | yes | yes | yes |
| 54 | High | left | political | yes | yes | yes |
| 55 | medium | middle | news | yes | yes | yes |
| 56 | medium | middle | news | yes | yes | yes |
| 57 | High | left | political | yes | yes | yes |
| 58 | High | right | political | yes | no | yes |
| 59 | High | right | political | yes | no | no |
| 60 | medium | right | tourism | yes | no | yes |
| 61 | medium | right | tourism | yes | yes | yes |
| 62 | medium | left | news | yes | no | yes |
| 63 | High | left | impression | yes | yes | no |
| 64 | Low | right | news | no | yes | no |
| 65 | High | left | political | yes | yes | yes |
| 66 | medium | left | impression | yes | yes | yes |
| 67 | medium | left | political | yes | yes | yes |
| 68 | High | right | political | yes | yes | yes |
| 69 | medium | left | political | yes | yes | yes |
| 70 | High | right | impression | yes | yes | no |
| 71 | medium | left | political | yes | yes | yes |
| 72 | medium | left | news | yes | yes | yes |
| 73 | medium | right | political | no | yes | no |
| 74 | Low | left | tourism | yes | no | no |
| 75 | High | left | news | yes | yes | yes |
| 76 | High | left | political | yes | yes | yes |
| 77 | Low | right | impression | no | no | yes |
| 78 | High | right | political | yes | yes | yes |
| 79 | High | left | impression | no | no | yes |
| 80 | medium | left | scientific | yes | yes | no |
| 81 | High | right | political | yes | yes | yes |
| 82 | Low | left | scientific | yes | yes | no |
| 83 | medium | right | tourism | yes | yes | no |
| 84 | Low | right | political | yes | yes | yes |
| 85 | high | left | impression | yes | no | yes |
| 86 | medium | left | tourism | yes | no | yes |
| 87 | medium | middle | impression | yes | no | yes |
| 88 | medium | middle | impression | no | yes | no |
| 89 | medium | right | scientific | yes | yes | no |
| 90 | medium | left | impression | no | no | yes |
| 91 | high | left | political | yes | yes | no |
| 92 | medium | left | news | no | yes | yes |
| 93 | high | left | political | yes | yes | yes |
| 94 | medium | right | news | yes | yes | no |
| 95 | medium | left | tourism | yes | no | yes |
| 96 | medium | middle | impression | yes | yes | yes |
| 97 | low | middle | impression | yes | no | no |
| 98 | low | right | impression | yes | no | no |
| 99 | medium | right | news | yes | yes | no |
| 100 | medium | left | impression | yes | yes | yes |

After providing database, we can pre-process the data and get primary information from them. It is identified that major community of bloggers belong to the political party of so-called reformists (leftists). This party has a great tendency to professional blogging. Among this, the party which is so-called traditional conservative (right-oriented) is in the next place that mostly has seasonal tendency to blogging. Although, bloggers without political orientation (moderate) has less numbers, they have professional approach to blogging (Diagram 1).

Most bloggers have bachelor degree, and then M.Sc. and Ph.D. graduates constitute this group which both has professional approach to blogging. Those with lower education don't have professional approach to blogging (Diagram 2). Those who believe in local political and social conditions effects on blogging and those who don't, have professional approach to blogging (Diagram 3). It is also the same about those who believe in local media function on the tendency toward blogging (Diagram 4).

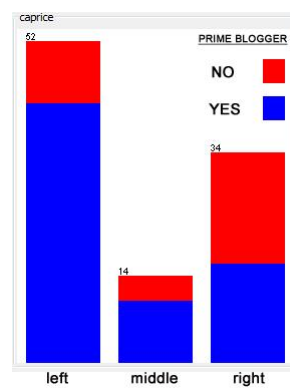
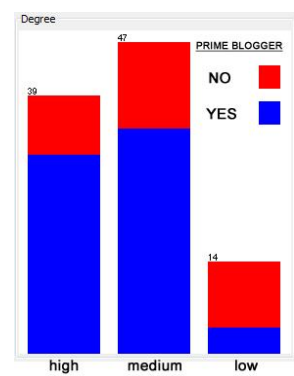

**Diagram 1: Abundance of Academic education and determining professional and seasonal tendency for each parameters**

**Diagram 2: Abundance of political tendencies and determining professional and seasonal tendency for each parameters**

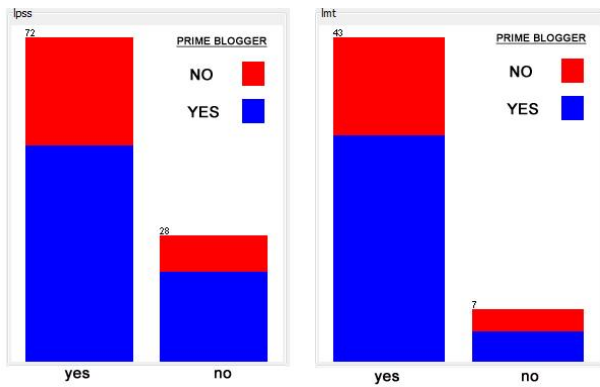

**Diagram 3: Abundance of local political and social conditions effects on blogging and determining professional and seasonal tendency for each parameters**

**Diagram 4: Abundance of local media function on the tendency toward blogging and determining professional and seasonal tendency for each parameters**

Among the interested subjects for blogging, politics has the first grade. Most people, who are interested in politics, are professional bloggers. The subjects such as personal, news and tourism are on the next grades. These also have professional approach. The only group who has professional and seasonal tendency to blogging is scientific bloggers (Diagram 5).

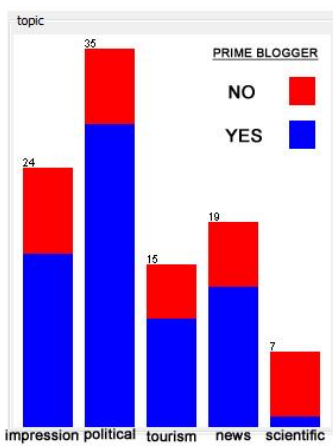

**Diagram 5: The abundance of interested subjects for digital writing and determining professional and seasonal tendency for each parameters**

After pre-processing of the data, we process them using weka3.6 tool and c4.5 algorithm. Then, we obtain decision tree which is shown in Figure 3.

```
topic = impression
| caprice = left
| | Degree = high
| | | lpss = yes: yes
| | | lpss = no: yes
| | Degree = medium: yes
| | Degree = low: null
| caprice = middle
| | Degree = high: null
| | Degree = medium
| | | lmt = yes: yes
| | | lmt = no: no
| | Degree = low: no
| caprice = right
| | lmt = yes: no
| | lmt = no: yes
topic = political
| lmt = yes
| | lpss = yes
| | | caprice = left
| | | | Degree = high: yes
| | | | Degree = medium: yes
| | | | Degree = low: null
| | | caprice = middle: null
| | | caprice = right: yes
| | lpss = no: yes
| lmt = no: no
topic = tourism
| Degree = high: yes
| Degree = medium
| | caprice = left: yes
| | caprice = middle: yes
| | caprice = right
| | | lpss = yes: no
| | | lpss = no: yes
| Degree = low: no
topic = news
| caprice = left: yes
| caprice = middle: yes
| caprice = right: no
topic = scientific
| caprice = left: no
| caprice = middle: yes
| caprice = right: no
```

**Figure 3- Decision Tree of the users' professional tendency based on different parameters effect on professional tendency**

By using decision tree (as shown in figure 3) and based on available data, it can be provided correct anticipation of bloggers behaviors and seen the role of each factors and their importance on professional approach. The error results and decision tree precision are presented in Table 2:

**Table 2: Error and precision rate in data classification**

| Class | ROC Area | F-Measure | Recall | Precision | FP Rate | TP Rate |
|---|---|---|---|---|---|---|
| Yes | 0.811 | 0.873 | 0.912 | 0.838 | 0.375 | 0.912 |
| No | 0.811 | 0.69 | 0.625 | 0.769 | 0.088 | 0.625 |

As a result of all these, we can say that there is a direct relationship between interesting subjects for writing and their tendency to blogging. The other results are reviewed by decision tree. It can be seen the results of decision tree in Table 3.

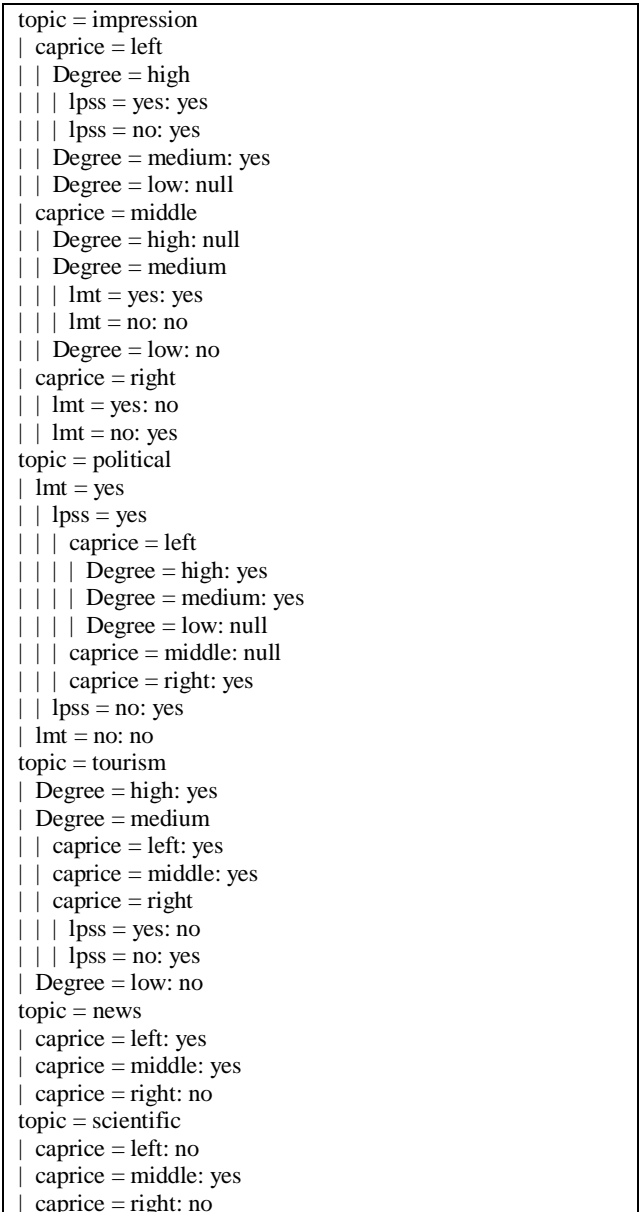

**Table 3: The Results of Decision Tree.**

| Accuracy | Num | Percent |
|---|---|---|
| Correctly Classified instances | 82 | 82 % |
| Incorrectly Classified Instances | 18 | 18 % |
| Kappa statistic | 0.5648 | - |
| K&B Relative Info Score | 5731.732 % | - |
| K&B Information Score | 51.5343 bits | 0.5153 bits/instance |
| Class complexity | order 0 | 90.5646 bits | 0.9056 bits/instance |
| Class complexity | Scheme | 11831.457 bits | 118.3146 bits/instance |
| Complexity improvement (SF) | -11740.8924 bits | -117.4089 bits/instance |
| Mean absolute error | 0.1834 | - |
| Root mean squared error | 0.3876 | - |
| Relative absolute error | 41.9651 % | - |
| Root relative squared error | 83.0247 % | - |
| Total Number of Instances | 100 | - |

So, determining, classifying parameters and making decision tree are provided with %82 precision. So, it can be used the above-mentioned results in decision makings and anticipations

## 6. CONCLOUSION AND FUTURE WORKS

In this paper, we propose a new approach for presenting the case study of bloggers tend to recognized parameters by using data mining. Due to performed simulations from input data of 100 users and bloggers of Kohkiloye and Boyer Ahmad Province and using weka3.6 tool and c4.5 algorithm to provide decision tree and achieve to future anticipation of users approach , results are shown with %82 precision.

If the users interested in writing memos, the basic factor of professional approach will be political thinking and academic education as the next step. If political issues are considered important in blogging, the attitude toward local media function, political and social conditions would be basic factors in recognizing professional approach. If the users interested in digital writing in tourism sector, academic education and political thinking would be basic factors in recognizing professional approach. And finally, if they choose subjects such as news and science, their professional approach would be political thinking.

The precision of decision tree in our paper and results have made us to provide strategic programs and software to planners based on decision tree in the future.

## 7. REFERENCES

[1] Zafarani,R, Jashki, M.A, Baghi,H.R , Ghorbani,A., 2008, A Novel Approach for Social Behavior Analysis of the Blogosphere, springer-Verlag Berlin Heidelberg, S. Bergler (Ed.): Canadian AI, 356–367.

[2] Wyld,D., 2007, The Blogging Revolution: Government in the Age of Web 2.0 ,IBM Center for The Business of Government.

[3] Juffinger,A., Lex, E., 2009, Cross language Blog Mining and Trend Visualization ,WWW 2009, 2009, Madrid, Spain.1149-1150.

[4] ManagemMacDougall, R. (2005), Identity, electronic ethos, and blogs: a technological analysis of symbolic exchange on the new news medium, American Behavioral Scientist, Vol. 49, No. 4,575–599.

[5] Demartini, G., Siersdorfer,S., Chelaru, S., Nejdl,W., Analyzing Political Trends in the Blogosphere, Proceedings of the Fifth International AAAI Conference on Weblogs and Social Media,465-469.

[6] Srinivasan R., Blog Analysis - Trends and Predictions, Applied Natural Language Processing Project Report,1-9.

[7] Moertini, V.S., 2003, Towards the Use of C4.5 Algorithm for classifying Banking Dataset , Oktober 2003, Integral, Vol. 8 No. 2. 105-117.

[8] Anyanwu, M.V., Shiva, S.G., Comparative Analysis of Serial Decision Tree Classification Algorithms, International Journal of Computer Science and Security, (IJCSS) Volume (3) : Issue (3). 230-240.

[9] Kumari,M, Godara, S, 2011, Comparative Study of Data Mining Classification Methodsin Cardiovascular Disease Prediction, IJCST , 2011,Vol. 2, Issue 2, 304-305.

[10] Han, J., Kamber, M., 2006, Data Mining: Concepts and TechniquesSecond Edition, Morgan Kaufmann Publishers.

[11] Gharehchopogh, F.S, 2010, Approach and Review of User Oriented Interactive Data Mining, the 4th International Conference on Application of Information and Communication Technologies (AICT2010), IEEE, Tashkent, Uzbekistan, 1-4.

[12] Daniel, T., 2005, Discovering knowledge in data, An Introduction to Data Mining, John Wiley & Sons, Inc., Hoboken, New Jersey.

[13] Gharehchopogh, F.S, Khalifelu, Z.A, 2011, Application Data Mining Methods for Detection Useful Knowledge in Health Center: A Case Study Using Decision Tree, International Conference on Computer Applications and Network Security (ICCANS 2011), 1-5.

[14] Lavanya, D., Usha R., 2011, Performance Evaluation of Decision Tree Classifiers on Medical Datasets, International Journal of Computer Applications (0975 – 8887), Volume 26– No.4, 1-4.

[15] Hartati,K, 2007, Implementation Of C4.5 Algorithm To Evaluate The Cancellation Possibility Of New Student Applicants At Stmik Amikom Yogyakarta, Proceedings of the International Conference on Electrical Engineering and Informatics Institute Technology Bandung, Indonesia June 17-19, 623-626

[16] Quinlan, J.R, 1986, Induction of Decision Trees, Machine Learning 1, Kluwer Academic Publishers, Boston. 81-106.

[17] Rosanna, E., Cassie, A. Bradley, E., Okdie, M, 2010, Personal Blogging Individual Differences and Motivations, IGI Global.292-301.